\documentclass[english,aps, twocolumn, pre,superscriptaddress, notitlepage]{revtex4-1}
\usepackage[T1]{fontenc}
\usepackage[latin9]{inputenc}
\usepackage{color}
\usepackage{mathtools}
\usepackage{amsmath}
\usepackage{amssymb}
\usepackage{graphicx}

\makeatletter

\providecommand{\tabularnewline}{\\}

\usepackage[colorlinks=true,linktocpage=true,linkcolor=MidnightBlue,urlcolor=MidnightBlue,citecolor=MidnightBlue,anchorcolor=MidnightBlue]{hyperref}
\usepackage[dvipsnames]{xcolor}
\usepackage{mhchem}
\renewcommand{\textendash}{--}

\makeatother

\usepackage{babel}
\begin{document}

\title{Relaxation to statistical equilibrium in stochastic Michaelis-Menten
kinetics}

\author{Subham Pal}

\affiliation{Department of Chemistry, Indian Institute of Technology, Madras,
Chennai-600036, India}

\author{Manmath Panigrahy}

\affiliation{Department of Chemistry, Indian Institute of Technology, Madras,
Chennai-600036, India}

\author{R. Adhikari}

\affiliation{DAMTP, Centre for Mathematical Sciences, University of Cambridge,
Wilberforce Road, Cambridge CB3 0WA, UK}

\author{Arti Dua}

\affiliation{Department of Chemistry, Indian Institute of Technology, Madras,
Chennai-600036, India}
\begin{abstract}
The equilibration of enzyme and complex concentrations in deterministic
Michaelis-Menten reaction networks underlies the hyperbolic dependence
between the input (substrates) and output (products). This relationship
was first obtained by Michaelis and Menten and then Briggs and Haldane
in two asymptotic limits: ``fast equilibrium'' and ``steady state''.
In stochastic Michaelis-Menten networks, relevant to catalysis at
single-molecule and mesoscopic concentrations, the classical analysis
cannot be directly applied due to molecular discreteness and fluctuations.
Instead, as we show here, such networks require a more subtle asymptotic
analysis based on the decomposition of the network into reversible
and irreversible sub-networks and the exact solution of the chemical
master equation (CME). The reversible and irreversible sub-networks
reach detailed balance and stationarity, respectively, through a relaxation
phase that we characterise in detail through several new statistical
measures. Since stochastic enzyme kinetics encompasses the single-molecule,
mesoscopic and thermodynamic limits, our work provides a broader molecular
viewpoint of the classical results, in much the same manner that statistical
mechanics provides a broader understanding of thermodynamics.
\end{abstract}
\maketitle

\section{Introduction\label{sec:Introduction}}

Michaelis and Menten, in their pioneering study of 1913, proposed
the initial rate method to estimate the kinetic parameters of enzymatic
reactions \cite{key-1}. They rationalized the experimentally observed
hyperbolic relationship between substrate concentration (the input)
and the enzyme velocity (the output) in terms of a simple mechanism
where enzyme $E$ and substrate $S$ instantaneously attained equilibrium
with the complex $ES$, characterized by an equilibrium constant $K_{M}$.
The input-output characteristic was parametrized by $K_{\text{M}}$,
now known as the Michaelis constant, and $V_{\text{max}}$, the maximum
attainable enzymatic velocity to give the celebrated hyperbolic relationship
\cite{key-2,key-3,key-4,key-5},

\begin{equation}
V_{0}=\frac{{V_{\text{max}}[S]}}{[S]+K_{M}}.\label{eq:IRV}
\end{equation}

The assumption that enzymes only weakly bound to the substrate, together
with the law of mass conservation, implied that a hyperbolic characteristic
curve would be attained when $[E]_{0}$, the initial enzyme concentration,
was much smaller than the substrate concentration. For over a century,
this has delineated the limit in which the initial rate method can
be reliably used in experimental assays to estimate kinetic parameters
\cite{key-2,key-3,key-4,key-5}.

Given the importance of the method established by Michaelis and Menten,
it is not surprising that it has been subject to refinement and scrutiny
since then \cite{key-6,key-7,key-8,key-9,key-10,key-11,key-12,key-13,key-14,key-15,key-16,key-17,key-18,key-19,key-20}.
In 1925, Briggs and Haldane introduced a more detailed mechanism 

\begin{equation}
E+S\xrightleftharpoons[k_{-1}]{k_{1}}ES\xrightarrow{^{k_{2}}}P+E\label{eq:MMM}
\end{equation}
in which the dissociation of the complex into product and the regeneration
of the enzyme was explicitly included and the assumption of instantaneous
equilibrium was relaxed \cite{key-7}. This allowed, for the first
time, the possibility of studying the initial transient in which equilibrium
between enzyme and substrate was yet to be established. In a subtle
conceptual shift, Briggs and Haldane derived an expression for the
steady-state velocity, beyond this initial transient, which, surprisingly,
had exactly the same analytic form as the Michaelis-Menten equation,
Eq. (\ref{eq:IRV}), but now with an explicit expression for the Michaelis
constant $K_{\text{M}}=\frac{k_{2}+k_{-1}}{k_{1}}$, in terms of the
rates of the elementary steps of the mechanism. Their analysis established
that equilibrium would be rapidly established if the dissociation
rate constant was much larger than the product formation rate constant
$k_{2}\ll k_{-1}$. Under this ``steady-state'' assumption, the initial
velocity $V_{0}$ in the Michaelis-Menten analysis could be identified
with the steady-state velocity $V_{ss}$ in the Briggs-Haldane analysis
if both association and dissociation rates were much larger than the
product formation rate. 

The conditions under which this identification can be made have been
refined further by Laidlar and remain an active area of research \cite{key-20,key-78,key-79,key-80,key-81,key-82,key-83,key-84,key-85},
see Ref. \cite{key-16} for a review. The characteristic curve remains
a central theoretical tool in both inferring reaction mechanisms from
experimental assays and in estimating the rate parameters of the inferred
mechanisms.

Over the last two decades, the increasing sophistication of experimental
methods has enabled the study of enzymatic catalysis at the single-molecule
level \cite{key-21,key-22,key-23,key-24,key-25,key-26,key-27,key-28,key-29,key-30,key-31,key-32}.
The thermodynamic limit implicit in the classical work described above
no longer applies to these experiments. Instead, the appropriate formalism
is that of continuous-time Markov processes with discrete state spaces,
the chemical master equation,reflecting the discrete nature of the
transition between catalytic states. The relaxation to statistical
equilibrium, consequently, is different from that in the thermodynamic
limit and presents, as we have discovered, several subtleties.

Bartholomay was the first to develop a chemical master equation (CME)
formalism for enzyme kinetics \cite{key-33}. This formalism incorporated
the randomness of discrete molecular collisions and their uncertainty,
termed molecular fluctuations, in each elementary step of the Michaelis-Menten
mechanism. Bartholomay used the CME to demonstrate the broader applicability
of the stochastic treatment and how number fluctuations influence
the mean catalytic response in the thermodynamic limit. Later works
used the quasi-steady-state assumption on the CME to reduce the two-dimensional
chemical master equation for the Michaelis-Menten (kinetic) mechanism,
Eq.(\ref{eq:MMM}), to a one-dimensional master equation for a Michaelis-Menten
(equilibrium) mechanism \cite{key-86,key-87,key-88}, see Ref. \cite{key-89}
for a review. However, this a priori assumption of stationarity cannot
fully characterise the molecular fluctuations arising from individual
reaction steps in the kinetic mechanism in the non-stationary transient
state, which is crucial for the asymptotic analysis of the classical
results.

We have developed a stochastic time-based approach (point process
description), which seamlessly combines with the number-based approach
(count process description), namely the chemical master equation,
and provides a comprehensive statistical analysis of molecular fluctuations
in enzymatic mechanisms \cite{key-34,key-35,key-36,key-37,key-38}.
This treatment encompasses the non-stationary transients and stationary
steady states in stochastic enzymatic networks, in time, and single-molecule,
mesoscopic and classical limits, in terms of the enzyme numbers.

In this work, we introduce new statistical measures for the count
and point process description and apply them to the exact solution
of the CME for the Michaelis-Menten mechanism to comprehensively analyse
the nature of molecular fluctuations in the transient and stationary
state kinetics. This provides a wide-ranging generalisation of the
asymptotic analysis of classical enzyme kinetics to stochastic enzyme
kinetics, with the subtle role played by molecular fluctuations being
clearly delineated. The novelty of our work lies in decomposing the
network of stochastic transitions into reversible and irreversible
subnetworks, where the reactions $E\rightarrow ES$ and $ES\rightarrow E$
correspond to the reversible subnetwork, and the reaction $ES\rightarrow E+P$
corresponds to the irreversible subnetwork. As we show, the reversible
network attains chemical detailed balance while the irreversible network
attains a stationary state through a non-trivial transient that we
carefully characterise. 

Sections \ref{sec:Deterministic-Michaelis-Menten-n} and \ref{sec:Stochastic-Michaelis-Menten-netw}
describe the salient features of the deterministic and stochastic
single-enzyme networks. After introducing the notations and terminology
of the count and point process descriptions comprising joint probabilities
of discrete species, generating function, waiting time distributions,
we develop the detailed statistical formalism in Sections \ref{sec:Generalized-rate-parameter}-\ref{sec:Results}.
A summary of each section is provided at the end of Section \ref{sec:Stochastic-Michaelis-Menten-netw}. 

\section{Deterministic Michaelis-Menten network \label{sec:Deterministic-Michaelis-Menten-n} }

\textcolor{black}{The deterministic Michaelis-Menten network is a
kinetic reaction scheme, Eq. (\ref{eq:MMM}), which involves a cyclical
process of enzyme turnovers through the catalytic rate constant $k_{2}$,
resulting in product formation over multiple cycles. Over time, the
rate of product formation reaches a stationary state, and the enzyme
turnover cycles reach an equilibrium state, where enzymes and their
complexes satisfy the chemical detailed balance condition, $k_{a}[E]_{eq}=k_{b}[ES]_{eq}$.
Here, $k_{a}=k_{1}[S]$ is the pseudo-first-order rate constant and
$k_{b}=k_{-1}+k_{2}$. This stationary (equilibrium) state underlies
the fundamental hyperbolic relation between the steady-state velocity
($V_{ss}$) and substrate concentration, quantified by the Michaelis-Menten
equation (MME), $V_{ss}=\frac{k_{2}k_{a}}{k_{a}+k_{b}}=\frac{k_{2}[E]_{0}[S]}{[S]+K_{M}}\equiv\frac{V_{max}[S]}{[S]+K_{M}}$,
where $[E]_{0}=[E]+[ES]$ is the total enzyme concentration, $V_{max}=k_{2}[E]_{0}$
is the maximum velocity at saturating substrate concentration, and
$K_{M}=(k_{-1}+k_{2})/k_{1}$.}

\textcolor{black}{The classical studies derive the MME using either
a fast-equilibrium assumption (FEA) or a quasi-steady-state assumption
(QSSA). The FEA assumes a time-scale separation between the fast equilibration
of enzymes and complexes and slow product formation. This assumption
implies a quasi-equilibrium between enzymes and complexes $\mathrm{E}\underset{k_{b}}{\stackrel{k_{a}}{\rightleftharpoons}}\mathrm{ES}$
\cite{key-1,key-9,key-10}. The chemical balance condition, thus obtained,
yields the MME, $V_{0}=\lim_{t\rightarrow0}\frac{d[P]}{dt}=k_{2}[ES]_{eq}=\frac{k_{2}[E]_{0}[S]}{[S]+K_{M}}$,
at the onset of the reaction. The QSSA does not assume an equilibrium
between enzymes and complexes but rather a short initial transient,
beyond which the rate of complex formation, $V_{c}=\frac{{d[ES]}}{dt}=k_{a}[E]-k_{b}[ES]$,
reaches a quasi-stationary state, where $V_{c}\approx0$ and $k_{a}[E]_{ss}=k_{b}[ES]_{ss}$.
This quasi-stationary state yields the MME, $V_{ss}=\lim_{t\rightarrow T^{*}}\frac{d[P]}{dt}=k_{2}[ES]_{ss}=\frac{k_{2}[E]_{0}[S]}{[S]+K_{M}}$
\cite{key-7,key-14,key-16}. }

\textcolor{black}{While the duration of the initial transient involves
non-stationary turnover kinetics, the FEA and SSA bypass this by assuming
a quasi-equilibrium or quasi-stationary state immediately after the
reaction begins. A simple estimate of the duration of the initial
transient follows from the solution of the rate equation for complex
in time, $V_{c}(t)=\frac{k_{a}[E]_{0}}{k_{a}+k_{b}}e^{-(k_{a}+k_{b})t}$.
For $t\gg T^{*}$, $V_{c}(t)$ relaxes to its stationary value, $[ES]_{ss}=\frac{k_{a}[E]_{0}}{k_{a}+k_{b}}$.
The QSSA, thus, translates into obtaining the duration of the initial
transient as }

\textcolor{black}{
\begin{equation}
T^{*}=\frac{{|\lambda|}}{(k_{a}+k_{b})}\label{eq:dtr}
\end{equation}
where $\lambda$ is a positive constant, which remains undetermined
in the mass action kinetics. For times beyond the initial transient
regime, the MME is exactly recovered. }

\textcolor{black}{The deterministic approach assumes enzymes and substrate
concentrations to be thermodynamically large. However, in vitro and
vivo, enzyme concentrations are significantly lower than the substrates.
This concentration ratio, $[E]_{0}\ll[S]$, is the basis of the initial
rate method and ensures that the initial rate of product formation
recovers the hyperbolic MME \cite{key-1,key-2,key-3,key-4,key-5}.
At such low enzyme concentrations, however, the turnover kinetics
is inherently stochastic, governed by molecular fluctuations. Whereas
the MM mechanism accounts for this concentration ratio through pseudo-first-order
rate constant $k_{a}=k_{1}[S]$, which implicitly assumes $[E]_{0}\ll[S]$,
the underlying mechanism and the rate equations only have a deterministic
context. They exclude the possibility of molecular fluctuations, which,
we show here, play a crucial role in describing the }\textcolor{black}{\emph{discrete}}\textcolor{black}{{}
turnover kinetics in the non-stationary transient regime and determining
its duration, Eq. (\ref{eq:dtr}).}

\textcolor{black}{In the next section, we describe the single-enzyme
stochastic MM network, in which the catalytic conversion of substrates
to products is carried out by a single enzyme, one substrate at a
time, in $p$ discrete turnover cycles. We examine the non-stationary
and stationary turnover kinetics of $N$ replicas of this stochastic
network from a probabilistic point of view. The stationarity and (statistical)
equilibrium conditions for the single or replica network follow from
the joint probability of the number of chemical species in the $p$-th
turnover cycle. These results provide a stochastic generalization
of the FEA, QSSA and their relation to the hyperbolic MME in the $p$-th
turnover cycle. }

\section{Stochastic Michaelis-Menten network\label{sec:Stochastic-Michaelis-Menten-netw}}

The stochastic description of the Michaelis-Menten network begins
by defining the state vector $\mathbf{n}=\{n_{\text{E}},n_{\text{ES}},n\}$,
and its joint probability $P(n_{\text{E}},n_{\text{ES}},n,t)$ at
time $t$, where $n_{\text{E}}$, $n_{\text{ES}},$ $n$ are the number
of enzymes, complexes, and products. The time evolution of the joint
probability obeys the Markovian chemical master equation (CME) \cite{key-33,key-39,key-40,key-41}:

\begin{eqnarray}
\frac{{\partial P(\mathbf{n},t)}}{\partial t} & = & k_{a}(n_{\text{E}}+1)P(n_{\text{E}}+1,n_{\text{ES}}-1,n;t)\nonumber \\
 & + & k_{-1}(n_{\text{ES}}+1)P(n_{\text{E}}-1,n_{\text{ES}}+1,,n;t)\nonumber \\
 & + & k_{2}(n_{\text{ES}}+1)P(n_{\text{E}}-1,n_{\text{ES}}+1,,n-1;t)\nonumber \\
 & - & [k_{a}n_{\text{E}}+(k_{2}+k_{-1})n_{\text{ES}}]P(\mathbf{n},t).\label{eq:cme-mm}
\end{eqnarray}
The CME accounts for the molecular discreteness and stochasticity
in each elementary step of the MM network. Since $N$ initial enzymes
either exist in free or bound state, the stochastic trajectories generated
from the CME obey the enzyme conservation law $n_{\text{E}}+n_{\text{ES}}=N$
at all times, implying $P(N-n_{\text{ES}},n_{\text{ES}},n,t)\equiv P(n_{\text{ES}},n,t|N)$.

\begin{figure}
\includegraphics[scale=0.54]{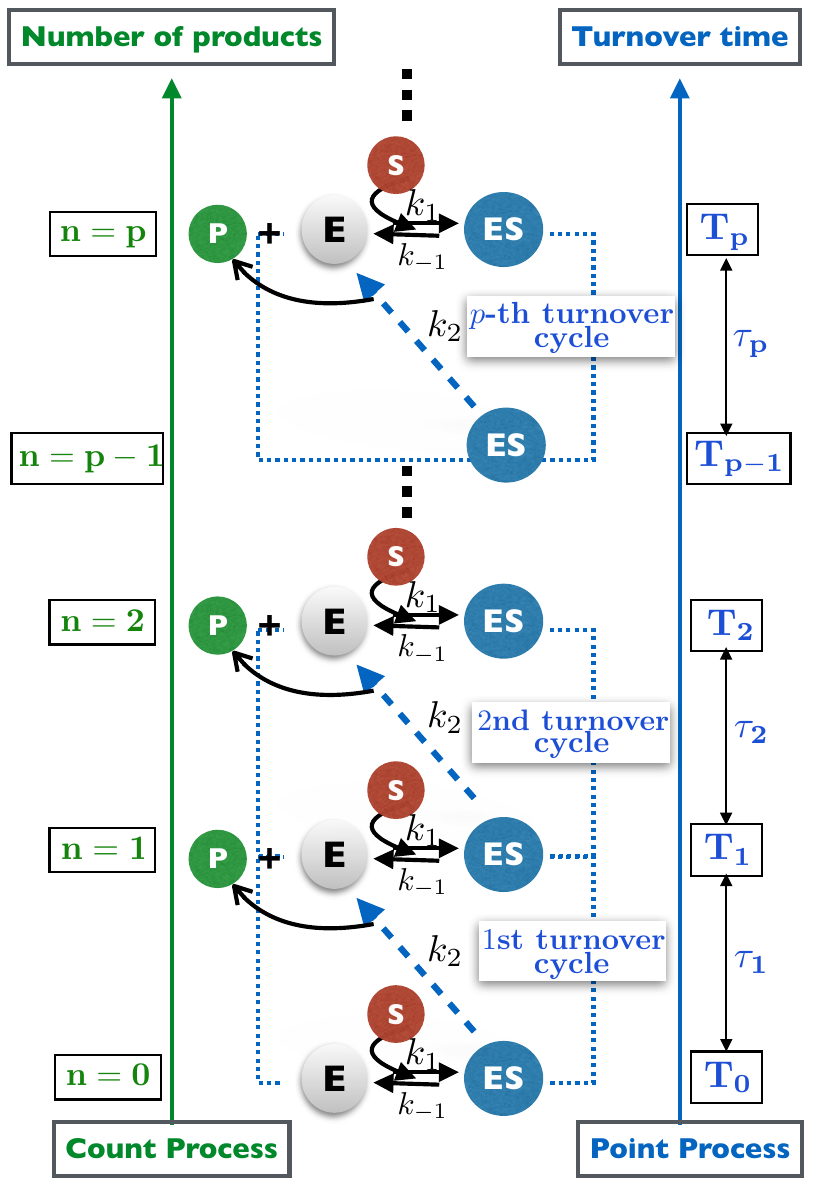}

\caption{A single-enzyme stochastic Michaelis-Menten (MM) network with discrete
turnover cycles $p=1,2,\cdots$, which include equilibrium between
$E$ and $ES$ (blue dotted lines), and discrete product turnover
events. Each turnover cycle includes $n=p-1$ number of products.
The turnover kinetics is characterized by the $p$-th waiting time
between two consecutive product turnovers, $\tau_{p}=T_{p}-T_{p-1}$,
where $T_{p}$ is the turnover time for the $p$-th product formation
starting from $T_{0}=0$ such that $\tau_{1}=T_{1}$. The product
turnovers are described in two complimentary ways: The count process
describes the number of products $n=p-1$ formed in continuous time
$t$; The point process description specifies the turnover time for
the $p$-th product formation. \label{fig:1}}
\end{figure}

Fig. (\ref{fig:1}) shows the salient features of a single-enzyme
stochastic Michaelis-Menten network with discrete turnover cycles
$p=1,2,\cdots$, which include equilibrium between $\text{E}$ and
$\text{ES}$ (blue dotted box), and discrete product turnovers indicated
by $n=p-1$. Products are formed one at a time, and their kinetics
involves three characteristic time scales: the turnover time for the
$p$-th product formation $T_{p}$ starting from $T_{0}=0$, the waiting
time between two consecutive product turnovers $\tau_{p}=T_{p}-T_{p-1}$,
and the lifetimes of $\text{E}$ and $\text{ES}$ states \cite{key-34,key-35,key-36,key-37}. 

In terms of the number of species vector $\mathbf{n}=\{n_{\text{E}},n_{\text{ES}},n\}$,
the statistical equilibrium between enzyme and complex in the $p$-th
turnover cycle, $[\text{E}\rightleftharpoons\text{ES}]_{p}$, corresponds
to reversible transitions between enzyme and complex, represented
by $\{1,0,p-1\}\rightleftharpoons\{0,1,p-1\}$. The irreversible catalytic
conversion, $\{0,1,p-1\}\rightarrow\{1,0,p\}$, marks the turnover
time $T_{p}$ for the $p$-th product formation, which is coupled
to enzyme renewal for the next cycle. The specification of the number
of products $n=p-1$ in time $t$ is the count process description,
described by the joint probability $P(n_{\text{ES}},n,t|N)$, which
is the solution of Eq. (\ref{eq:cme-mm}). The specification of the
turnover time $T_{p}$ for the $p$-th product is the point process
description, described in terms of the distributions of turnover times,
$w(T_{p}|N)$ \cite{key-42}.

The CME, Eq. (\ref{eq:cme-mm}), describes the time evolution of discrete
states $\mathbf{n}$ in continuous time $t$. It can be transformed
into a partial differential equation by using the following relation
between the generating function $G(s_{1},s_{2},t|N)$ and the joint
probability $P(n_{\text{ES}},n,t|N)$ \cite{key-34}: 
\begin{equation}
G(s_{1},s_{2},t|N)=\sum_{p}\sum_{n_{\text{ES}}}s_{1}^{n_{\text{ES}}}s_{2}^{p-1}P(n_{\text{ES}},p-1,t|N)\label{eq:genfunc1}
\end{equation}

\textcolor{black}{where $n_{\text{ES}}$ and $n=p-1$ are independent
variables. The transformation yields an equation of motion for $G(s_{1},s_{2},t|N)$,
which is continuous in the variables $s_{1}$ and $s_{2}$: }
\begin{eqnarray}
\frac{{\partial G(s_{1},s_{2},t|N)}}{\partial t} & = & k_{a}N(s_{1}-1)G(s_{1},s_{2},t|N)\nonumber \\
 & + & \left[(1-s_{1})(k_{b}+k_{a}s_{1})\right.\nonumber \\
 &  & \left.-k_{2}(1-s_{2})\right]\frac{\partial G(s_{1},s_{2},t|N)}{\partial s_{1}}\label{eq:Eq-of-motion}
\end{eqnarray}
\textcolor{black}{The exact solution of the partial differential equation
follows from the method of characteristic. The solution, presented
in our earlier work \cite{key-34}, is given by}

\textcolor{black}{
\begin{eqnarray}
G(s_{1},s_{2},t|N) & = & e^{-BNt}\left[\cosh(A^{\prime}t)\right.\nonumber \\
 &  & \text{\ensuremath{\left.+\frac{{k_{b}-k_{a}(1-2s_{1})}}{2A^{\prime}}\sinh(A^{\prime}t)\right]^{N}}}\label{eq:genfunc2}
\end{eqnarray}
where $B=\frac{(k_{a}+k_{b})}{2}$, $A^{\prime}=B\sqrt{{1-\delta(1-s_{2})}}$
and $\delta=\frac{4k_{a}k_{2}}{(k_{a}+k_{b})^{2}}$. }

The generating function $G(s_{1},s_{2},t|N)$ and its moments describe
the turnover kinetics in terms of the joint probability of the species
numbers at time $t$. The point process description complements the
latter by incorporating molecular details of stochasticity in the
substrate binding and unbinding times in the $p$-th cycle through
the distributions of turnover times, $w(T_{p}|N)$. The fundamental
relation between the two, derived in our earlier works \cite{key-34,key-35,key-36},
is given by
\begin{equation}
w(T_{p}|N)=\text{-\ensuremath{\sum_{n=0}^{p-1}\sum_{n_{ES}}\left.\frac{\partial P(n_{ES},n,t|N)}{\partial t}\right|_{t=Tp}}}\label{eq:count-point}
\end{equation}
\textcolor{black}{Eq. (\ref{eq:count-point}) holds for any $N$ and
$p$ and leads to the following exact expression for the turnover
time distributions in terms of the generating function:}

\textcolor{black}{
\begin{equation}
w(T_{p}|N)=\frac{{k_{2}}}{(p-1)!}[\partial_{s_{2}}^{p-1}\partial_{s_{1}}G(s_{1},s_{2},T_{p}|N)]_{s_{1}=1,s_{2}=0}.\label{eq:wtdp-1}
\end{equation}
}

In this work, we use the count and point process description of stochastic
Michaelis-Menten network, Eq. (\ref{eq:genfunc1})-(\ref{eq:wtdp-1}),
to address two fundamental questions in enzyme kinetics: Firstly,
what is the relationship between the experimental condition, $[E]_{0}\ll[S]$
widely used in the initial rate method, and the theoretical condition
$k_{2}\ll k_{-1}$ commonly used to analyze stationary (equilibrium)
states in kinetic mechanisms? Secondly, what is the duration of the
initial transient regime beyond which the enzymatic velocity attains
its stationary value and shows hyperbolic substrate dependence? 

The answers to these lie in the assumption made in the Michaelis-Menten
(MM) work, $[E]_{0}\ll[S]$, which is at odds with the deterministic
framework used to study their kinetics. At low enzyme concentrations,
the deterministic mechanism is replaced by a single-enzyme Michaelis-Menten
network, Fig. (\ref{fig:1}), and its replicas. This stochastic network
admits a more subtle stationary and equilibrium states described by
the joint probabilities, $P(n_{\text{ES}},p-1,T_{p})$, of discrete
chemical states ${\bf n}=(n_{\text{ES}},p-1)$, evaluated at times
$t=T_{p}$ with $n_{E}=N-n_{ES}$. 

In Section \ref{sec:Generalized-rate-parameter}, we derive explicit
expressions for the turnover number dependent joint probabilities.
We then obtain a static condition on the rate parameters of the Michaelis-Menten
network that ensures stationarity of the joint probabilities and leads
to a statistical equilibrium between the average number of enzymes
and complexes in the first turnover cycle. This condition is a stochastic
analogue of the fast-equilibrium assumption that describe the concentration-based
quasi-equilibrium in the Michaelis-Menten work. 

In Section \ref{sec:Duration-of-the}, we obtain a dynamic condition
on the rate parameters, which yields the duration of the transient
regime and provides a stochastic generalization of the quasi-steady-state
assumption in deterministic kinetics. 

Section \ref{sec:Turnover-number-dependent} relates the static and
dynamic conditions on the rate parameters with the steady-state velocity
in the first and $p$-th turnovers. The static and dynamic rate paramter
conditions are stochastic generalizations of the fast-equilibrium
and steady-state assumptions in deterministic kinetics that lead to
hyperbolic initial and steady-state velocities, respectively. 

In Section \ref{sec:Results}, we compare our theory with stochastic
simulations and summarize the key findings in Section \ref{sec:Summary-and-Conclusion}. 

\section{Stationarity, Statistical equilibrium and Generalized fast equilibrium
conditions \label{sec:Generalized-rate-parameter}}

We begin with the following axiom on the stationarity of the joint
probabilities, $P(n_{\text{ES}},p,t=T_{p+1})$, for successive enzyme
turnovers: 

\begin{eqnarray}
P(n_{\text{ES}},0,T_{1}) & =P(n_{\mathbf{\text{ES}}},1,T_{2}) & =P(n_{\text{ES}},2,T_{3})\nonumber \\
\cdots & =P(n_{\text{ES}},p,T_{p+1}) & =\cdots,\label{eq:stationarity-main}
\end{eqnarray}
where $p=1,2,\cdots$. 

For stationarity to be realized in the first turnover cycle $p=1$,
the joint probabilities of the first $P(n_{\text{ES}},0,T_{1})$ and
second $P(n_{ES},1,T_{2})$ turnover cycles in Eq. (\ref{eq:stationarity-main})
must be equal. The equality between the two naturally implies that
joint probabilities of all successive turnovers are equal. In this
notation, $P(n_{\text{ES}},p,T_{p+1})$ is the joint probability of
the stationary state, which is realized in the asymptotic limit of
a large number of products $p\rightarrow\infty$. 

In this section, we deduce a condition on the rate parameters of the
MM network that permits stationarity in the first turnover cycle.
Since this condition only involves rate parameters and not time, we
term it a static rate parameter condition (SRPC). This condition,
if obeyed, guarantees stationarity and statistical equilibrium between
enzymes and complexes in the first turnover cycle. While stationarity
condition compares joint probabilities of successive turnover cycles
$p=1,2,\cdots$, Eq. (\ref{eq:stationarity-main}), the condition
of statistical equilibrium follows from the joint distribution of
the $p$-th cycle. To highlight the difference, we introduce a new
kinetic measure, $R(T_{p})$, which relates the SRPC with the condition
of statistical equilibrium for the first turnover cycle $p=1$. Non-compliance
with the SRPC implies stationarity and statistical equilibrium are
not attained in the first turnover. In such cases, a dynamic rate
parameter condition is required, which we derive in the next section.
We conclude this section with a brief survey of how the SRPC provides
a stochastic generalization of the fast equilibrium assumption. 

The joint probabilities in Eq. (\ref{eq:stationarity-main}) can be
derived from the moment-generating function, Eq. (\ref{eq:genfunc1}).
The joint probabilities for $p=2,3,\cdots$ turnovers yield unwieldy
expressions as they involve partial derivatives with respect to $s_{2}$,
similar to Eq. (\ref{eq:wtdp-1}). We bypass their evaluation and
focus on the joint probabilities of two cases of interest: the joint
probability of the first turnover at $t=T_{1}$,

\begin{equation}
P(m,0,T_{1})=\frac{1}{m!}\left.\frac{\partial^{m}G(s_{1},s_{2},t)}{\partial s_{1}^{m}}\right|_{s_{1}=s_{2}=0}\label{eq:jpd1}
\end{equation}
and statioary state,

\begin{equation}
\lim_{p\rightarrow\infty}P(m,p,T_{p+1})=\frac{1}{m!}\left.\frac{\partial^{m}G(s_{1},s_{2},t)}{\partial s_{1}^{m}}\right|_{s_{1}=0,s_{2}=1}\label{eq:jpd2}
\end{equation}
where $m\equiv n_{\text{ES}}$. Substituting the explicit expression
for the generating function, Eq. (\ref{eq:genfunc2}), into Eqs. (\ref{eq:jpd1})
and (\ref{eq:jpd2}) yields the joint probability of the first turnover,

\begin{eqnarray}
P(m,0,T_{1})= & \left[^{N}C_{m}\left(\frac{k_{a}}{A}\right)^{m}\sinh^{m}(At)\left(\cosh(At)\right.\right.\nonumber \\
 & \left.\left.+\frac{(k_{b}-k_{a})}{2A}\sinh(At)\right)^{N-m}e^{-BNt}\right]_{t=T_{1}}\label{eq:probT1}
\end{eqnarray}
and stationary state,
\begin{eqnarray}
\lim_{p\rightarrow\infty}P(m,p,T_{p+1})= & \left[^{N}C_{m}(\gamma)^{N}(1-e^{-2Bt})^{m}\right.\nonumber \\
 & \times\left.\left(\frac{k_{b}}{k_{a}}+e^{-2Bt}\right)^{N-m}\right]_{t=T_{P+1}}\label{eq:probTp}
\end{eqnarray}
where $A=B\sqrt{1-\delta}$, $B=\frac{(k_{a}+k_{b})}{2}$ and $\gamma=\frac{{k_{a}}}{k_{a}+k_{b}}$. 

Comparison of Eqs. (\ref{eq:probT1}) and (\ref{eq:probTp}) shows
that $P(m,0,T_{1})$ is equivalent to $\lim_{p\rightarrow\infty}P(m,p,T_{p+1})$
in the asymptotic limit of $A^{N}\rightarrow B^{N}$ or $(1-\delta)^{N/2}\rightarrow1$.
To first order in $\delta$, it implies $N\delta\rightarrow0$ or
$N\delta\ll1$. From Eq. (\ref{eq:stationarity-main}) and aforementioned
analysis, it follows that the condition of stationarity for the first
turnover $(p=1)$ is given by

\begin{equation}
\lim_{N\delta\rightarrow0}P(n_{\text{ES}},0,T_{1})\equiv\lim_{p\rightarrow\infty}P(n_{\text{ES}},p,T_{p+1}).\label{eq:stationarity-1}
\end{equation}
It is to note that the asymptotic limit of $N\delta\rightarrow0$
is the SRPC condition, pertaining to $N\delta\ll1$. It provides a
stochastic generalization of the fast equilibrium assumption in deterministic
($N\rightarrow\infty$) kinetics. Below we analyze it for $N\geq1$. 

Single enzyme turnovers $(N=1)$, always satisfy the rate parameter
condition $\delta\ll1$ required for stationarity, Eq. (\ref{eq:stationarity-1}),
$\lim_{\delta\rightarrow0}P(1,0,T_{1})=P(1,p,T_{p+1})$ with $p=1,2,\cdots$.
To demonstrate this, we rewrite $\delta\ll1$ as $(k_{a}+k_{b})^{2}-4k_{a}k_{2}\gg0$.
Substituting $k_{b}=k_{-1}+k_{2}$ into the latter, we obtain $(k_{a}-k_{2})^{2}+k_{-1}^{2}+2k_{-1}(k_{a}+k_{2})\gg0$.
The latter reveals that $\delta\ll1$ is a positive definite, implying
that single-enzyme turnovers are stationary for all $p$. Another
interesting limit emerges by replacing $k_{a}$ with $k_{1}[S]$ and
$\frac{k_{b}}{k_{1}}$ with $K_{M}$. This permits us to rewrite the
inequality as $([S]+K_{M})^{2}\gg4N[S]k_{2}/k_{1}$, where $N=1$.
At the inflection point, when $K_{M}=[S]$, the SRPC for $N=1$ simplifies
to $k_{2}\ll\frac{k_{-1}}{(N-1)}$. This is a remarkable result as
it reveals that for $N=1$, the values of $k_{2}$ are unbound. It
underpins the renewal characteristics of single-enzyme turnovers.
We return to this point in the next section.

For $N>1$, the SRPC corresponds to $N\delta\ll1$ or $(k_{a}+k_{b})^{2}\gg4Nk_{a}k_{2}$,
which can be re-expressed as

\begin{equation}
\left(1+\frac{{K_{M}}}{[S]}\right)^{2}\gg\frac{N}{[S]}K_{c}\label{eq:srpc}
\end{equation}
where $K_{C}=\frac{k_{2}}{k_{1}}$ is the Van Slyke-Cullen constant
\cite{key-6,key-15,key-16}. 

Does the compliance with the rate parameter condition for stationarity
in the first turnover, Eq. (\ref{eq:srpc}), lead to statistical equilibrium
between enzymes and complexes at $t=T_{1}$ ? To verify this, we define
the ratio of the average number of enzymes as complexes in the $p$-th
turnover 

\begin{equation}
R(T_{p})=\left.\frac{{\langle n_{\text{E}}(t)\rangle}}{\langle n_{\text{ES}}(t)\rangle}\right|_{t=T_{p}}\label{eq:seq1}
\end{equation}
as the kinetic measure of statistical equilibrium. This ratio asymptotes
to its equilibrium value $\overline{R}_{p}$ at long times,

\begin{equation}
\overline{R}_{p}\equiv\left[\frac{\langle n_{\text{E}}\rangle_{\text{eq}}}{\langle n_{\text{ES}}\rangle_{\text{eq}}}\right]_{p}=\frac{k_{b}}{k_{a}}\label{eq:seq2}
\end{equation}
yielding the statistical equilibrium condition, $\left[k_{a}\langle n_{\text{E}}\rangle_{eq}=k_{b}\langle n_{\text{ES}}\rangle_{eq}\right]_{p}$,
in the $p$-th turnover cycle. 

Let us now analyze how Eqs. (\ref{eq:stationarity-main})-(\ref{eq:seq2})
are related to each other. Eq. (\ref{eq:stationarity-main}) is the
most general representation of the stationarity condition for $p$
successive turnovers. Eq. (\ref{eq:probT1}) is the (non-stationary)
joint probability of enzymes and complexes in the first turnover time
$T_{1}$; Eq. (\ref{eq:probTp}) represents the (stationary) joint
probability at $T_{p+1}$, attained asymptotically in the limit $p\rightarrow\infty$.
The left-hand side of Eq. (\ref{eq:stationarity-1}) signifies that
stationarity can be realized in the first turnover provided the rate
parameters of the MM network obey the SRPC, Eq. (\ref{eq:srpc}),
corresponding to $N\delta\ll1$. In the latter limit, $A\equiv B$
and the expression for the joint probability in Eq. (\ref{eq:probT1})
reduces to 

\begin{eqnarray*}
 & P(m,0,T_{1})\equiv\left[^{N}C_{m}\left(\frac{k_{a}}{B}\right)^{m}\left(\frac{1}{2^{N}}\right)(1-e^{-2Bt})^{m}\right.\\
 & \times\left.\left((1+e^{-2Bt})+\frac{k_{b}-k_{a}}{k_{b}+k_{a}}(1-e^{-2Bt})\right)^{N-m}\right]_{t=T_{1}}
\end{eqnarray*}
Upon further simplification, it leads to the following expression, 

\begin{eqnarray}
P(n_{\text{ES}},0,T_{1})\equiv & \left[^{N}C_{n_{\text{ES}}}(\gamma)^{N}(1-e^{-2Bt})^{n_{\text{ES}}}\right.\nonumber \\
 & \times\left.\left(\frac{k_{b}}{k_{a}}+e^{-2Bt}\right)^{N-n_{\text{ES}}}\right]_{t=T_{1}}\label{eq:PesT1}
\end{eqnarray}
which is equivalent to Eq. (\ref{eq:probTp}), \emph{i.e.}, the right-hand
side of Eq. (\ref{eq:stationarity-1}) with $m\equiv n_{ES}$. Crucially,
thus, the attainment of stationarity in the first turnover time depends
on compliance with the SRPC, Eq. (\ref{eq:srpc}).

For $N>1$, the first moments of Eq. (\ref{eq:PesT1}) are given by
$\langle n_{\text{E}}(t)\rangle_{t=T_{1}}=N\gamma\left(\frac{k_{b}}{k_{a}}+e^{-2BT_{1}}\right)$
and$\langle n_{\text{ES}}\rangle_{t=T_{1}}=N\gamma(1-e^{-2BT_{1}})$.
Substituting them into Eq. (\ref{eq:seq1}) yields $R(T_{1})=\frac{(k_{b}/k_{a}+e^{-2BT_{1}})}{(1-e^{-2BT_{1}})}$,
which asymptotes to $\overline{R}_{1}\equiv\left[\frac{\langle n_{\text{E}}\rangle_{\text{eq }}}{\langle n_{\text{ES}}\rangle_{\text{eq}}}\right]_{p=1}=\frac{k_{b}}{k_{a}}$
at $T_{1}\gg(k_{a}+k_{b})^{-1}$, recovering the statistical equilibrium
condition $\left[k_{a}\langle n_{\text{E}}\rangle_{eq}=k_{b}\langle n_{\text{ES}}\rangle_{eq}\right]_{p=1}$
in the first turnover. For $N=1$, the probability of enzyme and complex
states are equal to their average values: $P_{\text{E}}(T_{1})\equiv P(0,0,T_{1})=\langle n_{E}(t)\rangle_{t=T_{1}}$
and $P_{\text{E}S}(T_{1})\equiv P(1,0,T_{1})=\langle n_{ES}(t)\rangle_{t=T_{1}}$.
The statistical equilibrium condition for $N=1$ is thus $\left[k_{a}\overline{P}_{\text{E}}=k_{b}\overline{P}_{\text{ES}}\right]$,
which is valid for all turnovers. Here, $\overline{P}_{\text{E}}$
and $\overline{P}_{\text{ES}}$ are equilibrium probabilities. 

Eq. (\ref{eq:srpc}) provides a lower bound on the magnitude of $K_{c}$
for a specific ratio of $N$ to $[S]$. At the inflection point, $[S]=K_{M}$,
the above inequality can be analyzed in two alternative ways: Firstly,
when $[S]$ replaces $K_{M}$, it reduces to $k_{2}\ll\frac{k_{a}}{N}$
or $\frac{N}{[S]}\ll\frac{1}{K_{c}}$. While the former is the condition
on the rate parameters of the kinetic mechanism, the latter is the
experimental condition that assigns a lower limit to the ratio $N/[S]$.
Secondly, when $K_{M}$ replaces $[S]$ then $k_{2}\ll k_{-1}$. 

Eq. (\ref{eq:srpc}), in essence, is a stochastic generalization of
the fast-equilibrium assumption in deterministic kinetics and applies
to any $N$. In classical deterministic limit, $N\rightarrow\infty$,
the condition $k_{2}\ll\frac{k_{a}}{N}$ pertains to $k_{2}\rightarrow0$.
It implies quasi-equilibrium between enzymes and complexes at the
onset of the reaction. This limit was implicitly assumed in the work
of Michaelis and Menten and forms the basis of the initial rate method.
The condition $k_{2}\ll k_{-1}$, first appeared in the theoretical
work of Briggs and Haldane \cite{key-7}, explains how the kinetic
MM network can yield quasi-equilibrium between enzymes and complexes
at initial times. 

It is worthwhile to highlight the link between the conditions of stationarity,
Eq. (\ref{eq:stationarity-1}), and statistical equilibrium, Eq. (\ref{eq:seq2}),
in the first turnover with the SRPC, Eq. (\ref{eq:srpc}). If the
rate parameters of the MM network satisfy the SRPC, then the stationary
state is realized in the first turnover and the corresponding statistical
equilibrium equation for $p=1$ is obtained. If the SRPC is violated,
then stationary equilibrium state is not attained in the first turnover
cycle. The next section describes how this state is attained dynamically. 

\section{Relaxation to statistical equilibrium and generalized steady-state
condition \label{sec:Duration-of-the}}

Eq. (\ref{eq:probTp}) evaluates the joint probability of the stationary
state in the asymptotic limit of $p\rightarrow\infty$. An enzymatic
network, however, attains this limit in finite time, \emph{i.e.},
for times beyond an initial transient regime, $T_{p\gg p^{*}}$. This
notation combines the count and point process description in the following
manner: $T^{*}\equiv T_{p^{*}}$ is the critical turnover time that
defines the duration of the transient regime, and $p^{*}$ is the
corresponding critical turnover number beyond which the product numbers
attain their asymptotic value. For times beyond $T^{*}$, the turnover
kinetics relaxes from a non-stationary $t\ll T^{*}$ to a stationary
equilibrium $t\gg T^{*}$ state. This is a count process description,
which occurs in continuous time. Similarly, for turnovers beyond $p^{*}$,
the turnover statistics changes from non-renewal $p\ll p^{*}$ to
renewal $p\gg p^{*}$. This is the point process description in discrete
turnover numbers \cite{key-34,key-35,key-36,key-37}. The combination
of the two, $T_{p\gg p^{*}}$, specifies the time beyond which the
$p$-th turnover cycle attains stationarity. The stationarity condition,
thus, reads as

\begin{equation}
P(n_{\text{ES}},p-1,T_{p})=P(n_{\text{ES}},p,T_{p+1}),\quad(p\gg p^{*}).\label{eq:dynstationarity}
\end{equation}
Below, we derive an approximate expression for $p^{*}$ to show that
the condition $p\gg p^{*}$ is the dynamic equivalent of Eq. (\ref{eq:srpc}).
For this, we obtain $T^{*}$ from the count process description and
$\langle T_{p^{*}}\rangle$ from the point process description. The
equality between the two yields $p^{*}$ for short transients. We
discuss its relevance to the dynamic rate parameter condition (DRPC)
and Eq. (\ref{eq:dynstationarity}). We conclude this section with
a brief description of how $p^{*}$ can be evaluated from stochastic
simulations. 

Eq. (\ref{eq:dtr}) quantifies the duration of the classical transient,,
$T^{*}=\frac{|\lambda|}{(k_{a}+k_{b})}$, where $\lambda$ is a positive
constant that remains undetermined in the mass action kinetics. To
estimate $T^{*}$ and $\lambda$ in the classical limit, let us assume
that the stationarity condition, Eq. (\ref{eq:dynstationarity}),
is satisfied beyond the first turnover cycle, $\text{p\ensuremath{\gg}1}$.
This assumption implies that the turnover kinetics in the first turnover
cycle is non-stationary, governed by Eq. (\ref{eq:probT1}). The first
moment of Eq. (\ref{eq:probT1}) yields the average number of complexes
in the non-stationary state, $\langle n_{\text{ES}}(t)\rangle_{t=T_{1}}=\frac{Nk_{a}}{2A}\left[e^{-(B-A)t}-e^{-(B+A)t}\right]_{t=T_{1}}$.
The QSSA on this average number, $\left.\frac{d\langle n_{\text{ES}}(t)\rangle}{dt}\right|_{t=T^{*}}=0$
yields the duration of the initial transient, 

\begin{equation}
T^{*}=\frac{{(1-\delta)^{-1/2}}}{(k_{a}+k_{b})}\ln\text{\ensuremath{\left[\frac{1+(1-\delta)^{1/2}}{1-(1-\delta)^{1/2}}\right]}.}\label{eq:criticalt}
\end{equation}
Comparing Eqs. (\ref{eq:dtr}) and (\ref{eq:criticalt}) yields an
explicit expression for $\lambda$

\begin{equation}
\lambda=(1-\delta)^{-1/2}\ln\text{\ensuremath{\left[\frac{1+(1-\delta)^{1/2}}{1-(1-\delta)^{1/2}}\right]}},\label{eq:lambda-1}
\end{equation}
where $\delta$ lies between zero and one, $0<\delta<1$. It, thus,
violates the stationarity, Eqs. (\ref{eq:stationarity-1}), and statistical
equilibrium, (\ref{eq:seq2}), conditions in the first turnover. 

\begin{table*}
\centering%
\begin{tabular}{|c|c|c||c|c||c|c|}
\hline 
Case & Primary Condition & $\Theta_{1}^{(2)}$ & Secondary condition (a)  & $\Theta_{1}^{(2)}$ & Secondary condition (b)  & $\Theta_{1}^{(2)}$\tabularnewline
\hline 
\hline 
I & $[S]\ll K_{M}$  & $\frac{K_{M}^{2}}{K_{c}[S]+K_{M}^{2}}$ & $K_{c}[S]\ll K_{M}^{2}$  & 1 &  $K_{c}[S]\gg K_{M}^{2}$  & $\frac{K_{M}^{2}}{K_{c}[S]}$\tabularnewline
\hline 
II & $[S]\approx K_{M}$ & $\frac{4K_{M}}{K_{c}+4K_{M}}$ &  $K_{c}\ll4K_{M}$  & 1 &  $K_{c}\gg4K_{M}$  & $\frac{4K_{M}}{K_{c}}$\tabularnewline
\hline 
III & $[S]\gg K_{M}$ & $\frac{[S]}{K_{c}+[S]}$ &  $K_{c}\ll[S]$  & 1 &  $K_{c}\gg[S]$  & $\frac{[S]}{K_{c}}$\tabularnewline
\hline 
\end{tabular}\caption{The variation of $\Theta_{1}^{(2)}$ as a function of $[S]$, Eq.
(\ref{eq:theta1-N2}), for three limiting cases. Primary conditions
correspond to low, intermediate and high substrate concentrations.
Secondary conditions (a) and (b) correspond to low and high $K_{c}$,
respectively.}
\label{tab:lim-cases}
\end{table*}

The point process description determines the critical turnover number
and the duration of the initial transient, statistically, in terms
of the mean $p$-th turnover $\langle T_{p}^{(N)}\rangle$ and waiting
$\langle\tau_{p}^{(N)}\rangle$ times \cite{key-34,key-35,key-36}.
The mean $p$-th turnover time is the sum of the waiting times between
two consecutive turnovers $\langle T_{p}^{(N)}\rangle=\sum_{k=1}^{p}\langle\tau_{k}^{(N)}\rangle$,
where $\langle\tau_{p}^{(N)}\rangle=\langle T_{p}^{(N)}\rangle-\langle T_{p}^{(N)}\rangle$.
While $\langle T_{p}^{(N)}\rangle$ is evaluated from the first moment
of the turnover time distributions $w(T_{p}|N)$, Eqs. (\ref{eq:count-point})
and (\ref{eq:wtdp-1}), $\langle T_{p}^{(N)}\rangle=\int_{0}^{\infty}dT_{p}T_{p}w(T_{p}|N)$,
the waiting time distribution $w(\tau_{p})$ and its mean $\langle\tau_{p}^{(N)}\rangle$
follow from stochastic simulations \cite{key-43,key-44,key-45,key-46}.
These temporal distributions show remarkably different statistics
for $N=1$ and $N>1$. 

For $N=1$, the waiting time distributions are independent and identical
$w(\tau_{p}^{(1)})=w(\tau)$ for all $p$ \cite{key-34,key-35,key-36}.
This implies that the first turnover time distribution $w(T_{1}^{(1)})$
is identical to $w(\tau)$. From Eqs. (\ref{eq:count-point}) and
(\ref{eq:wtdp-1}), the exact expression for $w(\tau)$ is obtained
as $w(\tau)=\frac{k_{2}k_{a}}{2A}\left[e^{-(B-A)\tau}-e^{-(B+A)\tau}\right]$.
The mean of $w(\tau)$ yields $\langle\tau\rangle=\frac{(k_{a}+k_{b})}{k_{2}k_{a}}$,
which is identical for $p=1,2,\cdots$. This feature characterizes
the renewal turnover statistics for $N=1$. It emerges from the absence
of the waiting time correlations between consecutive turnovers for
$N=1$, $C_{1q}^{(\text{N})}=\langle\delta\tau_{1}^{\text{(N)}}\delta\tau_{1+q}^{\text{(N)}}\rangle=0$,
where $\delta\tau_{q}^{(\text{N)}}=\tau_{q}^{(\text{N)}}-\langle\tau_{q}^{(\text{N})}\rangle$
and $q=1,2,\cdots$. This leads to two interesting results: first,
the mean $p$-th turnover time follows the renewal theorem $\langle T_{p}^{(1)}\rangle=\sum_{k=1}^{p}\langle\tau_{k}\rangle=p\langle\tau\rangle$;
second, the inverse of $\langle\tau\rangle$ yields the single-enzyme
velocity in the first turnover $v_{1}=\langle\tau\rangle^{-1}=\frac{k_{a}k_{2}}{k_{a}+k_{b}}$.
The renewal turnover statistics implies that the interval between
two consecutive turnovers is the same $\langle\tau\rangle$ for $p=1,2,\cdots.$
It is worth noting that two important results of the previous section
for $N=1$, $\delta\ll1$ and $k_{2}\ll\frac{k_{-1}}{(N-1)}$, are
linked to the renewal nature of the single enzyme turnovers. All these
results imply that single enzyme turnovers satisfy the stationarity
condition, Eq. (\ref{eq:stationarity-1}), at $p=1$ and do not have
a transient regime. 

For $N>1$, the waiting time distributions are non-identical $w(\tau_{p}^{(N)})\neq w(\tau_{p+1}^{(N)})$
for turnovers below $p^{*}$, but become identical $w(\tau_{p}^{(N)})=w(\tau_{p+1}^{(N)})$
in the asymptotic limit of $p\gg p^{*}$ \cite{key-34,key-35,key-36}.
This effect emerges from the presence of non-stationary waiting time
correlations $C_{1q}^{(\text{N})}\neq0$ which persist for $q=1,2,\cdots p^{*}$
and vanish $C_{1q}^{(\text{N})}=0$ beyond $p^{*}$. The critical
turnover number, thus, demarcates the non-renewal regime ($p\ll p^{*}$)
where the turnover statistics is non-stationary from the renewal regime
($p\gg p^{*}$) where the stationarity is obtained asymptotically.
In the stationary state, the mean waiting times between successive
products turnover $\langle\tau_{p\gg p^{*}}\rangle$ are identical
$\langle\tau_{p\gg p^{*}}\rangle=\langle\tau\rangle/N$. In the non-renewal
regime, these times are non-identical and depend on the turnover number. 

In the point process description, the duration of the transient is
defined by the mean critical turnover time $\langle T_{p^{*}}^{\text{(N)}}\rangle=\sum_{k=1}^{p^{*}}\langle\tau_{k}^{\text{(N)}}\rangle$.
In the classical limit, where the kinetics of all but the first turnover
is non-stationary $\langle T_{p^{*}}^{\text{(N)}}\rangle=\sum_{k=1}^{p^{*}}\langle\tau_{k}^{\text{(N)}}\rangle=\langle\tau_{1}^{(N)}\rangle+(p^{*}-1)\langle\tau\rangle/N$.
For short transients, it is reasonable to assume $\langle\tau_{1}^{(N)}\rangle\approx\langle\tau\rangle/N$.
This assumption of renewal statistics allows us to write $\langle T_{p^{*}}^{\text{(N)}}\rangle\approx p^{*}\langle\tau\rangle/N$.
The count process description of this duration is given by Eq. (\ref{eq:criticalt}).
Combining the results of count and point processes, $T^{*}\approx\langle T_{p^{*}}^{\text{\text{(N)}}}\rangle\approx p^{*}\langle\tau\rangle/N$,
yields an approximate analytical expression for $p^{*}\approx T^{*}N/\langle\tau\rangle\approx|\lambda|N\delta$,
which simplifies to

\begin{equation}
p_{\text{theor}}^{*}\approx|\lambda|\frac{{N[S]K_{c}}}{([S]+K_{M})^{2}}.\label{eq:crit-turno}
\end{equation}
As stated above, $p\gg p^{*}$ is the dynamic rate parameter condition
(DRPC) that marks the critical turnover beyond which the stationarity
condition, Eq. (\ref{eq:dynstationarity}), is obeyed. From Eq. (\ref{eq:crit-turno}),
it follows that the DRPC is given by 

\begin{equation}
p\gg|\lambda|\frac{{N[S]K_{c}}}{([S]+K_{M})^{2}}\label{eq:drpc}
\end{equation}
The DRPC subsumes the SRPC, Eq. (\ref{eq:srpc}), and provides a stochastic
generalization of the QSSA. The rate parameters that satisfy Eq. (\ref{eq:srpc})
yield $p^{*}\approx0$, implying that the stationary, Eq. (\ref{eq:stationarity-1}),
and statistical equilibrium, Eq. (\ref{eq:seq2}), conditions are
obeyed in the first turnover cycle. The rate parameters that disobey
Eq. (\ref{eq:srpc}) yield $p\gg p^{*}$, where $p^{*}$ given by
Eq. (\ref{eq:crit-turno}). For $p\gg p^{*}$, thus, the stationarity
and statistical equilibrium conditions, Eq. (\ref{eq:dynstationarity})
and Eq. (\ref{eq:seq2}), are realized, dynamically, in the $p$-th
turnover cycle. 

The theoretical estimate for critical turnover number, Eq. (\ref{eq:crit-turno}),
assumes renewal turnover statistics and thus provides a lower bound
on $p^{*}$. The latter is valid for short transients, \emph{i.e.},
for rate parameters that weakly violate the SRPC. For long transients,
the product turnovers follow non-renewal statistics characterized
by non-stationary waiting time correlations, $C_{1q}^{(\text{N})}=\langle\delta\tau_{1}^{\text{(N)}}\delta\tau_{1+q}^{\text{(N)}}\rangle$.
Thus, the critical turnover number beyond which $C_{1p^{*}}=0$ provides
a numerical estimate for $p_{\text{num}}^{*}$. We obtain the latter
from stochastic simulations and present the results in Section \ref{sec:Results}. 

\section{Kinetic measure of hyperbolicity \label{sec:Turnover-number-dependent}}

Eqs. (\ref{eq:srpc}) and (\ref{eq:drpc}) provide the static and
dynamic rate parameter conditions for stationarity and statistical
equilibrium in the first and $p$-th turnover cycles, abbreviated
as the SRPC and DRPC, respectively. To link these conditions with
the hyperbolic MME, we introduce a dimensionless kinetic measure -
the turnover number dependent fractional enzymatic velocity (FEV),

\begin{equation}
\Theta_{p}^{\text{(N)}}=\frac{{V_{p}^{\text{(N)}}}}{V_{\text{ss}}^{\text{(N)}}}\label{eq:thetap-1}
\end{equation}
where $V_{p}^{(N)}$ is the point process description of the enzymatic
velocity in the $p$-th turnover cycle and $V_{ss}^{(N)}=\frac{k_{2}N[S]}{[S]+K_{M}}$
is the classical steady-state velocity. The FEV, Eq. (\ref{eq:thetap-1}),
as a ratio of the two velocities, provides a single parameter to quantify
the deviation of $V_{P}$ from hyperbolic velocity $V_{ss}$ for $p=1,2,\cdots$.
In this form, it seamlessly combines two crucial features of the discrete
turnover kinetics: first, that the $V_{p}^{(N)}$ is non-hyperbolic
in the non-renewal (or non-stationary) transient regime with $V_{p\ll p^{*}}^{(N)}\neq V_{ss}^{(N)}$,
implying $\Theta_{p\ll p^{*}}\neq1$; second, that $V_{p}$ attains
stationarity in the renewal steady-state, \emph{i.e.}, $V_{p\gg p^{*}}^{(N)}=V_{ss}^{(N)}$,
implying $\Theta_{p\gg p^{*}}=1$. 

Both these features are included in the $p$-th enzymatic velocity,
whose functional form $V_{p}^{(N)}=p\langle T_{p}^{(N)}\rangle^{-1}$
incorporates the renewal turnover statistics, $\langle T_{p}^{(N)}\rangle=p\langle\tau^{(N)}\rangle=p\langle\tau\rangle/N$,
in the steady state \cite{key-34,key-35,key-36,key-37}. The renewal
statistics defines the single-enzyme velocity $v_{1}=\langle\tau\rangle^{-1}=\frac{p}{\langle T_{p}^{(1)}\rangle}$
for all $p$. As stated earlier, for $N=1$, the renewal turnover
statistics, $\langle T_{p}^{(1)}\rangle=p\langle\tau\rangle$, yields
the single-enzyme velocity $v_{1}=\langle\tau\rangle^{-1}=\frac{k_{a}k_{2}}{k_{a}+k_{b}}$.
This is the single-enzyme analogue of the classical MME, $v_{1}=\frac{V_{ss}^{(N)}}{N}=\frac{{k_{2}[S]}}{[S]+K_{M}}$,
and represents the hyperbolic relation between $v_{1}$ and $[S]$
in the first turnover. For $N>1$, the turnover statistics is non-renewal
$\langle T_{p\ll p^{*}}^{(N)}\rangle\neq p\langle\tau\rangle/N$ in
the transient state and renewal $\langle T_{p\ll p^{*}}^{(N)}\rangle=p\langle\tau\rangle/N$
in the steady-state . This leads to deviation from the MME in the
transient state, $V_{p\ll p^{*}}^{(N)}\neq V_{ss}^{(N)}$, and its
asymptotic recovery in the steady-state, $V_{p\gg p^{*}}=p\langle T_{p\gg p^{*}}^{\text{(N)}}\rangle^{-1}=N\langle\tau\rangle^{-1}=V_{\text{ss}}^{\text{(N)}}$. 

\begin{figure*}
\centering\includegraphics[clip,width=0.8\textwidth]{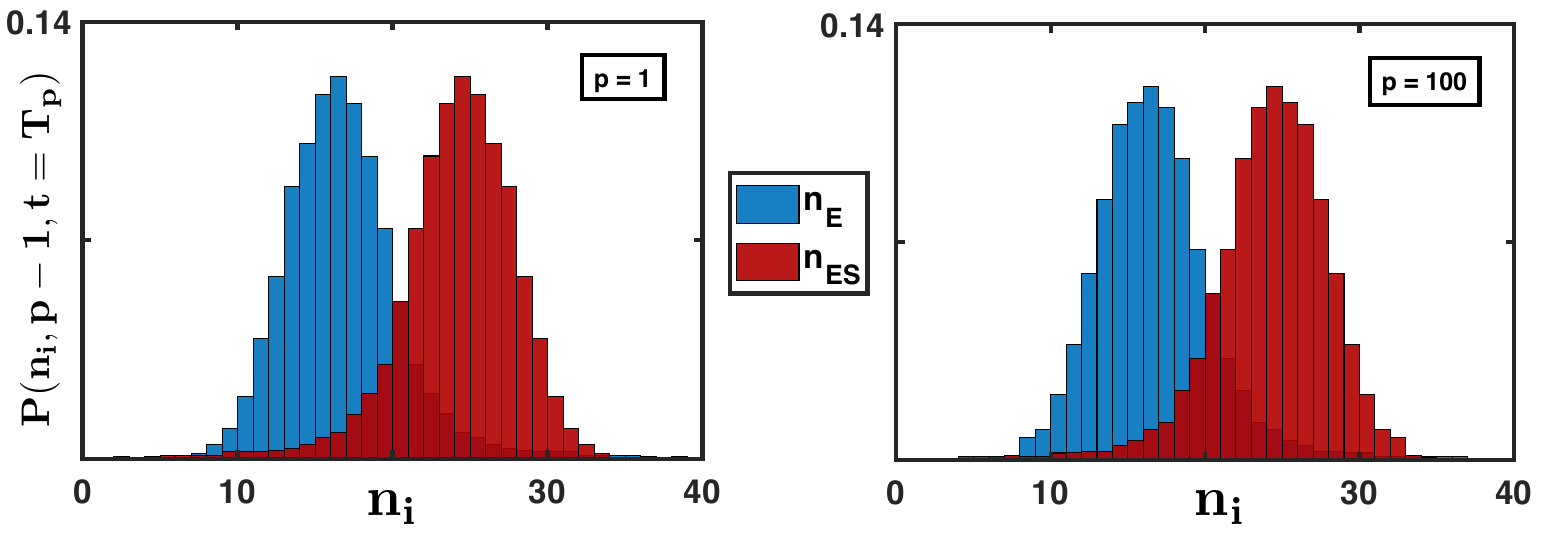}

\caption{Joint probabilities of the number of enzymes ($n_{\text{E}}$), complexes
($n_{\text{ES}}$) and products ($n=p-1$) at time $t=T_{p}$, $P(n_{\mathbf{\text{i}}},p-1,T_{p})$,
with $\text{\textbf{i}}=\{\text{E},\text{ES}\}$. These distributions
are obtained from stochastic simulations for the rate parameter values,
$[S]=2.5$, $K_{M}=1.5$, $N=40$, $K_{c}=0.01$. These parameters
obey the SRPC, Eq. (\ref{eq:srpc}), and hence satisfy the condition
of stationarity, Eq. (\ref{eq:stationarity-1}), leading to the equivalence
between distributions at $p=1$ and $p=100$. The first moments of
these distributions yield the statistical averages of $\text{ES}$
and $\text{E}$ as $\langle n_{\text{ES}}\rangle_{p=1}=23.3$ and
$\langle n_{\text{E}}\rangle_{p=1}=16.3$. These estimates are close
to their equilibrium values $\langle n_{\text{ES}}\rangle_{eq}=\frac{{N[S]}}{[S]+K_{M}}=25$
and $\langle n_{\text{E}}\rangle_{eq}=N-\langle n_{\text{ES}}\rangle_{eq}=15$.\label{fig:2}}
\end{figure*}

\begin{figure*}
\centering\includegraphics[clip,width=0.8\textwidth]{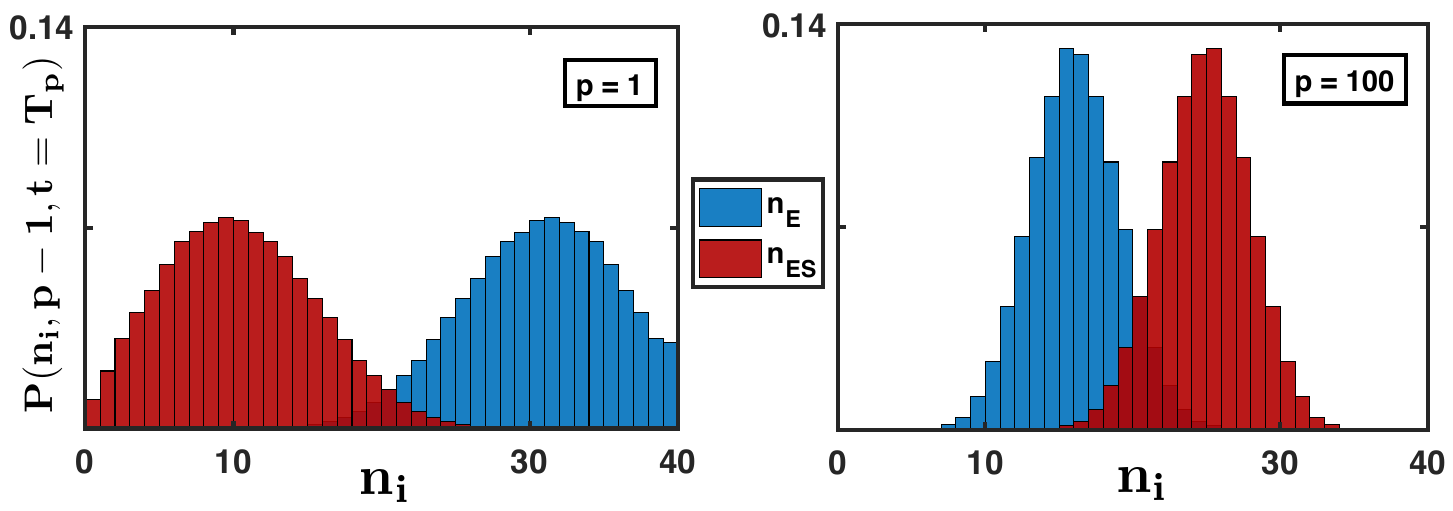}

\caption{Joint probability distributions, $P(n_{\mathbf{\text{i}}},p-1,t=T_{p})$,
for parameter values $[S]=2.5$, $K_{M}=1.5$, $N=40$, $K_{c}=1$.
These rate parameter values violate the SRPC, Eq. (\ref{eq:srpc}),
and hence do not satisfy the stationarity condition, Eq. (\ref{eq:stationarity-1}),
at $p=1$. For these rate parameters, the equilibrium averages are
$\langle n_{\text{E}}\rangle_{eq}=15$ and $\langle n_{\text{ES}}\rangle_{eq}=25$;
and the critical turnover number, obtained from Eq. (\ref{eq:crit-turno}),
is $p_{\text{theor }}^{*}=15$. The first moment of the distributions
at $t=T_{1}$ yields the average number of $\text{E}$ and $\text{ES}$
as $\langle n_{\text{E}}\rangle_{p=1}=30\protect\neq\langle n_{\text{E}}\rangle_{eq}$
and $\langle n_{\text{ES}}\rangle_{p=1}=10\protect\neq\langle n_{\text{ES}}\rangle_{eq}$,
which deviate from the stationary (equilibrium) values. For $p=100$
($p\gg p^{*}$), the first moments of the joint distributions yield
number averages $\langle n_{\text{E}}\rangle_{p=100}=15.6$ and $\langle n_{\text{ES}}\rangle_{p=100}=24.4$,
which are close to their equilibrium estimates. \label{fig:3}}
\end{figure*}

In essence, the $p$-dependent FEV, Eq. (\ref{eq:thetap-1}), includes
waiting time correlations between successive turnovers, and quantifies
the degree of non-hyperbolicity, \emph{i.e.}, deviation from the MME,
in the transient regime. The increase in $p$ brings about a crossover
from $\Theta_{p\ll p^{*}}^{(N)}\neq1$ in the non-stationary state
to $\Theta_{p\gg p^{*}}^{(N)}=1$ in the stationary state. The asymptotic
recovery of the MME, $\Theta_{p\gg p^{*}}=1$, in the stationary state
is linked to the DRPC. Below, we obtain an exact expression for the
substrate dependence of $\Theta_{1}^{(2)}$. We deduce rate parameter
conditions that yield $\Theta_{1}^{(2)}\neq1$ and show how the limit
$\Theta_{1}^{(2)}\rightarrow1$ is linked to the SRPC. We generalize
these results in the next section. 

For $N=2$ and $p=1$, the FEV is given by $\Theta_{1}^{(\text{2})}=V_{1}^{(\text{2})}/V_{\text{ss}}$,
where $V_{1}=\langle T_{1}^{\text{(2)}}\rangle^{-1}$. The first moment
of $w(T_{1}|2)$, Eq. (\ref{eq:wtdp-1}), evaluated in Mathematica,
yields $V_{1}^{\text{(2)}}=\langle T_{1}^{\text{(2)}}\rangle^{-1}=\frac{2k_{2}[S]([S]+K_{M})}{K_{c}[S]+([S]+K_{M})^{2}}$
and the FEV as 

\begin{equation}
\Theta_{1}^{\text{(2)}}=\frac{([S]+K_{M})^{2}}{K_{c}[S]+([S]+K_{M})^{2}}.\label{eq:theta1-N2}
\end{equation}
To understand the link between Eq. (\ref{eq:theta1-N2}) and the SRPC,
we consider the variation of $\Theta_{1}^{\text{(2)}}$ as a function
of $[S]$ for three limiting cases, summarized in Table \ref{tab:lim-cases}. 

\begin{figure*}
\centering\includegraphics[clip,width=1\textwidth]{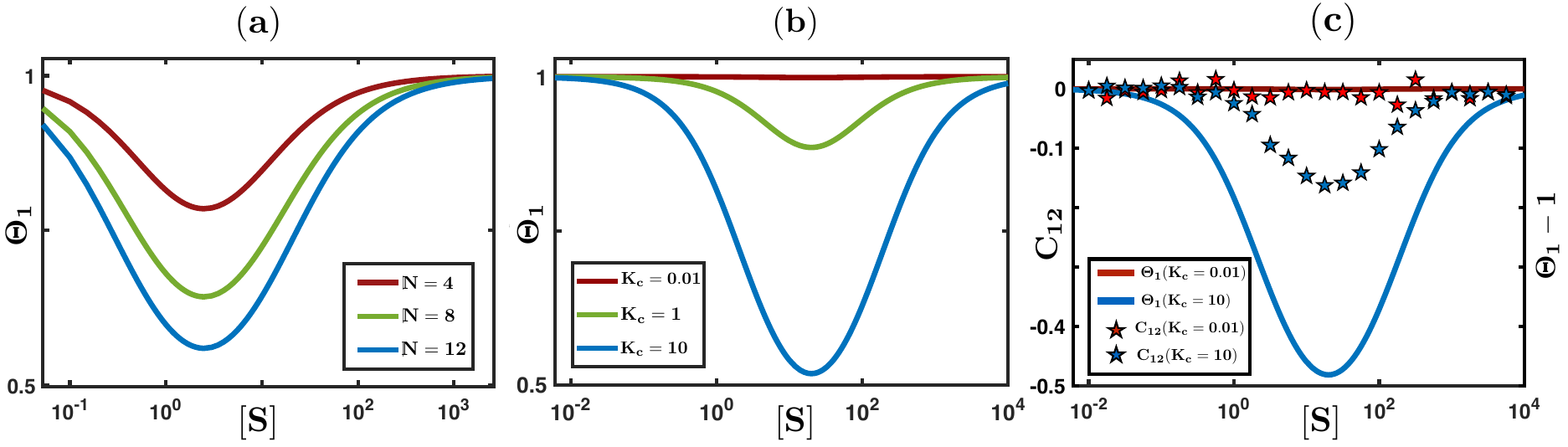}

\caption{Variation of $\Theta_{1}$ as a function of substrate concentration
$[S]$ (a) for a range of $N$ at $K_{M}=2.5$, $K_{c}=1$; (b) for
a range of $K_{c}$ at $K_{M}=20$ and $N=12$. The relation between
the substrate variation of the waiting time correlations between first
and second turnovers, $C_{12}$, and scaled FEV, $(\Theta_{1}-1)$,
for the rate parameters of (b) is captured in (c). The U-shaped curves
represent non-hyperbolic substrate dependence of the enzymatic velocity,
$\Theta_{1}<1$, in the non-classical transient regime. The recovery
of the hyperbolic MME in the steady state is signaled by $\Theta=1$
for (a) and (b) and $C_{12}=(\Theta_{1}-1)=0$ for (c). \label{fig:4}}
\end{figure*}

\begin{figure*}
\centering\includegraphics[clip,width=1\textwidth]{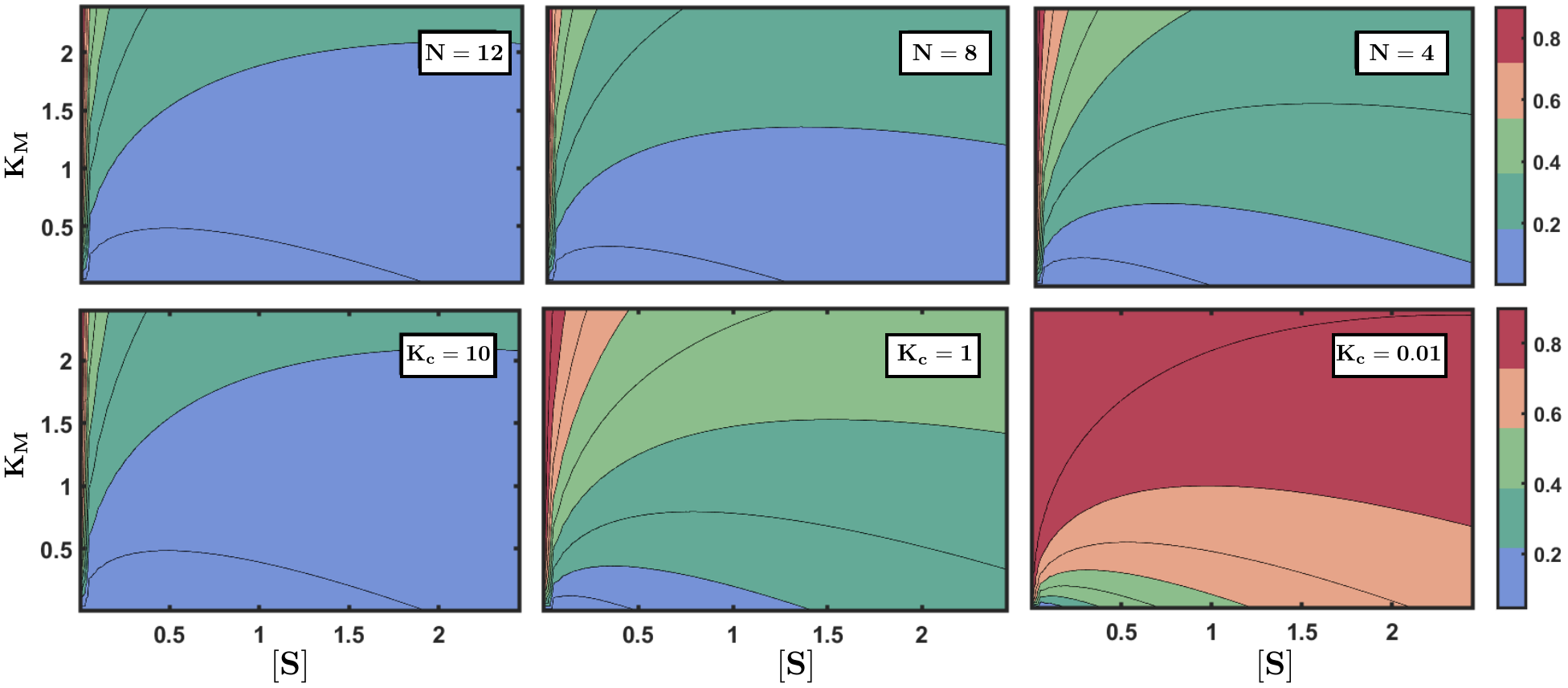}

\caption{Contour plots of $\Theta_{1}$, with $K_{M}$ and $[S]$ as independent
parameters, for a range of $N$ at $K_{c}=10$ (top panel) and a range
of $K_{c}$ at $N=10$ (bottom panel). From left to right, the non-classical
transient region, represented by the iso-values $\Theta_{1}<1$, reduces
in size with the decrease in $N$ and $K_{c}$. On the right, the
approach to the steady state is indicated by the iso-values close
to one, $\Theta_{1}\approx1$.\label{fig:5}}
\end{figure*}

At low $K_{c}$, the rate parameter conditions for low {[}case Ia{]},
intermediate {[}case IIa{]}, and high {[}case IIIa{]} substrate concentrations
suggest $\Theta_{1}^{\text{(2)}}\approx1$ for all $[S]$. The latter
implies that, at low $K_{c}$, the hyperbolic substrate dependence
of the MME is the result of statistical equilibrium, between $N$
independent and identical $\text{E}$ and $\text{ES}$ states, in
the first catalytic turnover cycle. At high $K_{c}$, in contrast,
the rate parameter conditions for low {[}case Ib{]}, intermediate
{[}case IIb{]}, and high {[}case IIIb{]} substrate concentrations
yield $\Theta_{1}^{\text{(2)}}<1$. For $[S]\ll K_{M}$, $\Theta_{1}^{\text{(2)}}$
decreases with increasing $[S]$; reaches a minimum at $[S]\approx K_{M}$;
for $[S]\gg K_{M}$, $\Theta_{1}^{\text{(2)}}$ increases with increasing
$[S]$ and asymptotically attains $\Theta_{1}^{\text{(2)}}\approx1$
when $K_{c}\ll[S]$ {[}case (3a){]}. 

The above analysis shows the SRPC, if obeyed, yields $\Theta_{1}^{(2)}=1$.
If violated yields $0<\Theta_{1}^{(2)}<1$. In the next section, we
generalize these results to show that the presence of the waiting
time correlations between first and successive turnovers, $C_{1q}^{(N)}\neq0$,
underlies the non-hyperbolic substrate response, $\Theta_{1}^{(N)}\neq1$,
in the transient regime. In the stationary state, vanishing of correlations,
$C_{1q}^{(N)}=0$ leads to hyperbolic substrate dependence of the
MME, $\Theta_{1}^{(N)}=1$. 

\begin{figure*}
\centering\includegraphics[clip,width=0.9\textwidth]{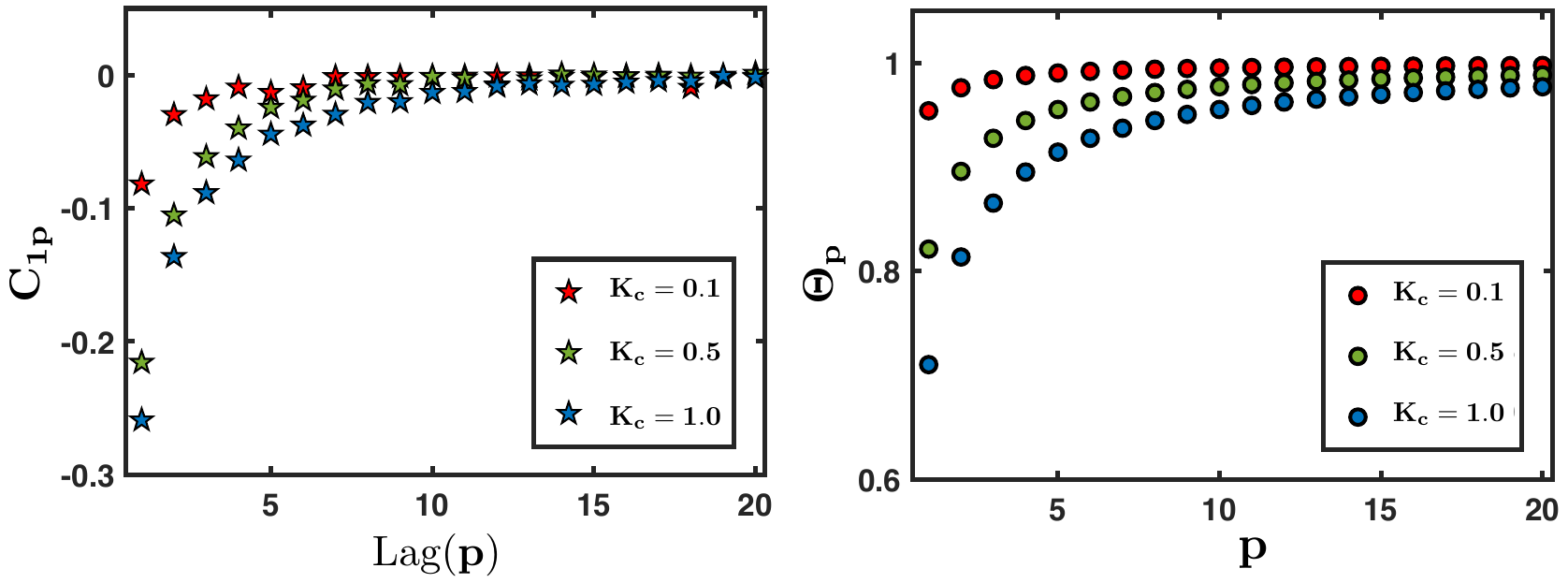}

\caption{The waiting time correlations $C_{1p}=\langle\delta\tau_{1}\delta\tau_{1+p}\rangle$
between the first and $(1+p)$ successive turnovers (left), and the
$p$-th dependent fractional enzyme velocity $\Theta_{p}$ (right)
as a function of $p$ for a range of $K_{c}$ at $[S]=2.5$, $K_{M}=1.5$,
$N=40$, obtained from the stochastic simulations. The critical turnover
number beyond which $C_{1p^{*}}=0$ yields the numerical estimates
$p_{\text{num }}^{*}=$ 4, 10, 18 at $K_{c}=$ 0.1, 0.5, 1, respectively.
These are close to the theoretical estimates $p_{\text{theor}}^{*}\approx$
2.7, 9, 14.6 from Eq. (\ref{eq:crit-turno}). \label{fig:6}}
\end{figure*}

\section{Comparison of theory with computer experiment\label{sec:Results}}

Equations (\ref{eq:stationarity-main}), (\ref{eq:srpc}), (\ref{eq:seq2}),
(\ref{eq:drpc}), and (\ref{eq:thetap-1}) are the key findings of
our theory and provide stochastic generalizations of the stationarity
condition, the fast-equilibrium assumption, the quasi-equilibrium
condition, the quasi-steady-state assumption, and the enzyme velocity,
respectively. Below, we combine the results of Sections \ref{sec:Generalized-rate-parameter}-\ref{sec:Turnover-number-dependent}
to establish that compliance with the SRPC, Eq. (\ref{eq:srpc}),
yields stationarity, statistical equilibrium and the MME in the first
turnover. Non-compliance implies that these conditions and the MME
are recovered dynamically for $p\gg p^{*}$, i.e. the DRPC, Eq. (\ref{eq:drpc}).

We begin our analysis by establishing the link between the SRPC, and
the DRPC with the stationarity Eq. (\ref{eq:stationarity-main}),
and statistical equilibrium, Eq. (\ref{eq:seq2}), conditions. For
this, we first obtain the joint probability distributions $P(n_{\text{E}},p,t=T_{p})$
and $P(n_{\text{ES}},p,t=T_{p})$ from the Doob-Gillespie stochastic
simulation algorithm \cite{key-43,key-44,key-45}. We generate $10^{6}$
stochastic trajectories and extract the joint distributions of the
number of enzymes $n_{\text{E}}$ and complexes $n_{\text{ES}}$ for
$n=p-1$ products at time $t=T_{p}$ \cite{key-46}. The simulation
results are shown in Figs. (\ref{fig:2}) and (\ref{fig:3}), which
we discuss below. 

Fig. (\ref{fig:2}) considers the rate parameter values (lower $K_{C}$)
that obey the SRPC in the first turnover cycle ($p=1$). The compliance
with the SRPC leads to the equivalence between the joint distributions
for $p=1$ and $p=100$ turnovers, $P(n_{\mathbf{\text{i}}},0,T_{1})\equiv P(n_{\mathbf{\text{i}}},p,T_{p+1})$,
for $p\gg p^{*}$ , where $\mathbf{i}=\{\text{E},\text{ES}\}$. The
number averages of $P(n_{\text{E}},0,T_{1})$ and $P(n_{\text{ES}},0,T_{1})$
yield $\langle n_{\text{E}}\rangle_{p=1}$ and $\langle n_{\text{ES}}\rangle_{p=1}$.
These averages satisfy the statistical equilibrium condition in the
first turnover cycle, $\left[k_{a}\langle n_{\text{E}}\rangle_{eq}=k_{b}\langle n_{\text{ES}}\rangle_{eq}\right]_{p=1}$. 

Fig. (\ref{fig:3}), in contrast, considers the rate parameter values
(higher $K_{c}$) that do not obey the SRPC. As a result, the conditions
of stationarity $P(n_{\mathbf{\text{i}}},0,T_{1})\neq P(n_{\mathbf{\text{i}}},p,T_{p+1})$
and statistical equilibrium $\left[k_{a}\langle n_{\text{E}}\rangle_{seq}\neq k_{b}\langle n_{\text{ES}}\rangle_{seq}\right]_{p=1}$
are not obeyed in the first turnover cycle. The DRPC yields $p_{\text{theor}}^{*}=15$.
For $p=100$, thus, the number averages of the stationary distribution
($p\gg p^{*}$), $P(n_{\mathbf{\text{i}}},p,T_{p+1})$, yield $\langle n_{\text{E}}\rangle_{p\gg p*}$
and $\langle n_{\text{ES}}\rangle_{p\gg p^{*}}$. These averages satisfy
the statistical equilibrium condition, $\left[k_{a}\langle n_{\text{E}}\rangle_{eq}=k_{b}\langle n_{\text{ES}}\rangle_{eq}\right]_{p\gg p}$
for $p\gg p^{*}$. 

A quantitative link between the SRPC and the substrate variation of
the first-FEV, $\Theta_{1}^{\text{(2)}}$, is established in Table
\ref{tab:lim-cases} for three limiting cases: $[S]\ll K_{M}$; $[S]\approx K_{M};$$[S]\gg K_{M}$.
Together they suggested that at high $K_{c}$, the violation of the
SRPC indicates non-hyperbolic enzymatic velocity in the transient
regime. For the latter, the substrate variation of $\Theta_{1}^{(\text{N})}$
yields a U-shaped curve, confined between $0$ and $1$, with a minimum
given by case IIb. 

\begin{table*}
\centering%
\begin{tabular}{|c|c|c||c|}
\hline 
Rate parameter condition & Stationarity condition & Statistical Equilibrium Condition & FEV \tabularnewline
\hline 
\hline 
SRPC obeyed & $P(n_{\text{\text{\textbf{i}}}},p-1,T_{p})=P(n_{\text{\textbf{\text{i}}}},p,T_{p+1})$,
$p=1,2,\cdots$ & $\left[k_{a}\langle n_{\text{E}}\rangle_{eq}=k_{b}\langle n_{\text{ES}}\rangle_{eq}\right]_{p=1}$ & $\Theta_{1}^{\text{(N)}}=1$\tabularnewline
\hline 
SRPC violated  & $P(n_{\mathbf{\text{\text{i}}}},p-1,T_{p})\neq P(n_{\text{\textbf{\text{i}}}},p,T_{p+1})$,
$p=1,2,\cdots$ & $\left[k_{a}\langle n_{\text{E}}\rangle_{eq}\neq k_{b}\langle n_{\text{ES}}\rangle_{eq}\right]_{p=1}$ & $\Theta_{1}^{\text{(N)}}<1$\tabularnewline
\hline 
DRPC obeyed & $P(n_{\mathbf{\text{\text{i}}}},p-1,T_{p})=P(n_{\text{\textbf{\text{i}}}},p,T_{p+1})$,
$p\gg p^{*}$ & $\left[k_{a}\langle n_{\text{E}}\rangle_{eq}=k_{b}\langle n_{\text{ES}}\rangle_{eq}\right]_{p\gg p^{*}}$ & $\Theta_{p\gg p^{*}}^{\text{(N)}}=1$\tabularnewline
\hline 
\end{tabular}\caption{Quantitative link between the static and dynamic rate parameter conditions,
Eqs. (\ref{eq:srpc}) and (\ref{eq:drpc}), with the stationarity
condition, Eq. (\ref{eq:stationarity-main}), the statistical equilibrium
condition, Eq. (\ref{eq:seq2}), and the $p$-th dependent fractional
enzyme velocity (FEV), $\Theta_{p}^{\text{(N)}}$, Eq. (\ref{eq:thetap-1}). }
\label{tab:comparison-1}
\end{table*}

Figs. (\ref{fig:4}a) and (\ref{fig:4}b) show the substrate dependence
of $\Theta_{1}^{(\text{N})}$, computed from Eq. (\ref{eq:thetap-1}),
for a range of $N$ and $K_{c}$, in Mathematica. For convenience,
all the rate parameters, including $K_{M}$, $K_{c}$ and $[S]$ are
dimensionless. For large $N$ and $K_{c}$, the SRPC is violated,
leading to U-shaped curves for the substrate variation of $\Theta_{1}^{(\text{N})}$
(Fig (\ref{fig:4}a)). With the decrease in the value of $K_{c}$,
when the SRPC is obeyed. $\Theta_{1}^{(\text{N})}=1$ for all $[S]$
(the red curve in Fig. (\ref{fig:4}b)). This limit recovers the hyperbolic
enzymatic velocity $V_{1}$, the MME, in the first turnover. The results
captured in Figs. (\ref{fig:4}a) and (\ref{fig:4}b) are consistent
with the analytical analysis for $N=2$, summarized in Table \ref{tab:lim-cases}. 

Fig. (\ref{fig:5}) represents the contour plots of $\Theta_{1}^{(\text{N})}$,
with $K_{M}$ and $[S]$ as independent variables, for a range of
$N$ at fixed $K_{c}$ (top panel), and a range of $K_{c}$ at fixed
$N$ (bottom panel). The iso-values lower than one, $\Theta_{1}<1$,
correspond to the non-stationary transient regime (blue and green
colors), and $\Theta_{1}\approx1$ (red color) correspond to the steady-state
regime, The top panel shows that the first turnover is in the transient
regime for larger $N$ and close to the steady-state for smaller $N$.
The bottom panel shows that the first catalytic turnover is in the
transient regime for higher $K_{c}$ and close to the steady-state
at lower $K_{c}$. Both these are in agreement with the predictions
of the SRPC, Eq. (\ref{eq:srpc}).

To trace the molecular origins of the U-shaped response curve, we
plot the substrate variation of the waiting time correlations between
first and second turnovers, $C_{12}^{(\text{N})}$, and scaled first-FEV,
$(\Theta_{1}^{(\text{N})}-1)$, Fig. (\ref{fig:4}c). The substrate
dependence of $C_{12}^{(\text{N})}$ is negative and shows a non-monotonic
response to increase in $[S]$ (blue stars). This is similar to the
U-shaped substrate variation of $(\Theta_{1}^{(\text{N})}-1)$ at
high $K_{c}$ (blue curve). This implies that at high $K_{c}$, when
the SRPC is violated, the first turnover is in the (non-renewal) transient
regime, governed by the waiting time correlations, \emph{i.e}., $C_{12}^{(N)}<0\implies\Theta_{1}^{(N)}<1$.
At low $K_{c}$, both $C_{12}^{(\text{N})}$ and $(\Theta_{1}^{(\text{N})}-1)$
are zero (red stars and curve). This implies that at low $K_{c}$,
when the SRPC is obeyed, the first turnover is in the (renewal) steady-state
regime, \emph{i.e.}, $C_{12}^{(N)}=0\implies\Theta_{1}^{(N)}=1$.
The inflection point for both $C_{12}^{(N)}$ and $\Theta_{1}^{(N)}$
is attained at $[S]\approx K_{M}$. 

The above analysis quantifies the link between the SRPC and FEV for
$p=1$. In particular, it illustrates that the SRPC, if violated,
yields $\Theta_{1}^{(\text{N})}<1$, and if obeyed, leads to $\Theta_{1}^{(\text{N})}=1$.
In the former case, the DRPC estimates the critical turnover number
$p^{*}$ beyond which the statistical equilibrium between $\text{E}$
and $\text{ES}$ is established and hyperbolicity is recovered. Eq.
(\ref{eq:crit-turno}) provides an analytical expression for the critical
turnover number $p_{\text{theor}}^{*}$, which is only valid for short
transients. To check the validity of the assumption made in deriving
Eq. (\ref{eq:crit-turno}), we compare the theoretical and numerical
estimates of $p^{*}$. Fig. (\ref{fig:6}a) shows the variation of
waiting time correlations $C_{1p}^{(\text{N})}$ with turnover lag
$p=1,2,\cdots$ obtained from stochastic simulations. The critical
turnover number beyond which $C_{1p^{*}}^{(\text{N})}=0$ provides
a numerical estimate for $p_{\text{num}}^{*}$. For short transients
(low $K_{c}$), the numerical estimate is very close to the theoretical
estimate. 

To quantify the link between the DRPC and the $p$-th dependent FEV,
we plot in Fig. (\ref{fig:6}b) the variation of $\Theta_{p}^{(\text{N})}$
as a function of $p$ for the same parameter as Fig. (\ref{fig:6}a).
The variation of $C_{1p}^{(\text{N})}$ versus $p$ is qualitatively
similar to $\Theta_{p}^{(\text{N})}$ versus $p$. In particular,
for $p\ll p^{*}$, the turnover kinetics in transient regime, where
negative waiting time correlations $C_{1p}^{(\text{N})}<0$ yield
non-hyperbolic response curves $\Theta_{p}^{(\text{N})}<1$, \emph{i.e.},
$C_{1p}^{(N)}<0\implies\Theta_{p}^{(\text{N})}<1$. For $p\gg p^{*}$,
$C_{1p}^{(\text{N})}=0$ yields the hyperbolic MME, $\Theta_{p\gg p^{*}}^{(\text{N})}=1$. 

Table (\ref{tab:comparison-1}) summarizes the link between the conditions
of statistical equilibrium and stationarity with the SRPC, DRPC and
the fractional enzymatic velocity. These results for mesoscopic enzyme
kinetics are valid for any number of enzymes $(N)$ and turnover number
($p$). 

\section{Summary and conclusion \label{sec:Summary-and-Conclusion}}

In this work, we present a novel statistical route to quantify, at
the molecular level, the number and temporal fluctuations in the stochastic
Michaelis-Menten network of $N$ enzymes as they form products in
discrete turnover times $T_{p}$. Our work begins by introducing a
condition of stationarity, which demands equality between the joint
probabilities of discrete species numbers in the $p$-th and $(p+1)$-th
cycles with $p=1,2,\dots$. We examine the statistical properties
of the joint probability through the count and point process descriptions
to show that the stationarity condition can be satisfied, statically
or dynamically, in the first or the $p$-th turnover cycle. 

We use the count process description to deduce the static condition
on the rate parameters of the network that yield statistical equilibrium
in the first turnover. We use the combination of the count and point
process descriptions to show that non-compliance with the static rate
parameter condition is a transient effect originating from the non-stationary,
or equivalently, the non-renewal turnover kinetics. We derive the
critical turnover number and time above which the turnover statistics
have a renewal character. This feature yields the dynamic rate parameter
condition that ensures the simultaneous realization of the stationarity
and statistical equilibrium conditions for $p\gg p^{*}$ turnovers.
It also provides an analytical estimate of the duration of the short
transients. 

We use the count process description to prove that single enzyme turnovers
in the stochastic Michaelis-Menten network, irrespective of the magnitude
of $k_{2}$, always obey the static and dynamic rate parameter conditions.
This feature corresponds to the renewal turnover statistics in the
point process description and implies that single enzyme turnovers
do not admit a transient regime \cite{key-36}. For a mesoscopic number
of enzymes, the static and dynamic rate parameter conditions provide
the stochastic generalization of the fast-equilibrium and steady-state
assumptions. 

In conclusion, our work traces the molecular origins of the widely
used fast-equilibrium and steady-state assumptions in enzyme kinetics
through the count and point process descriptions of turnover probability.
It combines the conditions of stationarity and statistical equilibrium
with hyperbolic reaction rate, providing stochastic generalizations
of the initial and steady-steady state velocity. The generalized theoretical
framework proposed in this study encompasses the classical (deterministic)
and single-enzyme (stochastic) Michaelis-Menten kinetics. The dynamic
rate parameter conditions, derived here, offer a novel way to quantify
the duration of the transient regime in mesoscopic Michaelis-Menten
kinetics through the statistical analysis of number and temporal fluctuations. 
\begin{acknowledgments}
MP acknowledges the financial support from the Council of Scientific
and Industrial Research (CSIR), Government of India.
\end{acknowledgments}

\end{document}